\documentstyle[12pt]{article}
\begin{document}
\newcommand{\beq}{\begin{equation}}
\newcommand{\eeq}{\end{equation}}
\newcommand{\beqa}{\begin{eqnarray}}
\newcommand{\eeqa}{\end{eqnarray}}
\newcommand{\barr}{\begin{array}}
\newcommand{\earr}{\end{array}}
\newcommand{\nonum}{\nonumber}
\thispagestyle{empty}
\begin{center}
\LARGE \tt \bf{Non-Riemannian geometry of vortex acoustics}
\end{center}
\vspace{2.5cm}
\begin{center} {\large L.C. Garcia de Andrade\footnote{Departamento de F\'{\i}sica Teorica - UERJ.

Rua S\~{a}o Fco. Xavier 524, Rio de Janeiro, RJ

Maracan\~{a}, CEP:20550-003 , Brasil.

E-mail.: garcia@dft.if.uerj.br}}
\end{center}
\vspace{1.0cm}
\begin{abstract}
The concept of acoustic metric introduced previously by Unruh (PRL-1981) is extended to include Cartan torsion by analogy with the scalar wave equation in Riemann-Cartan (RC) spacetime. This equation describes irrotational perturbations in rotational non-relativistic fluids. This physical motivation allows us to show that the acoustic line element can be conformally mapped to the line element of a stationary torsion loop in non-Riemannian gravity. Two examples of such sonic analogues are given. The first is when we choose the static torsion loop in teleparallel gravity. In this case Cartan torsion vector in the far from the vortex approximation is shown to be proportional to the quantum vortex number of the superfluid. Also in this case the torsion vector is shown to be proportional to the superfluid vorticity in the presence of vortices. Torsion loops in RC spacetime does not favor the formation of superfluid vortices. It is suggested that the teleparallel model may help to find a model for superfluid neutron stars vortices based on non-Riemannian gravity. 
\end{abstract}
\vspace{0.5cm}
\vspace{2.0cm}
\newpage
\pagestyle{myheadings}
\markright{\underline{Non-Riemannian geometry of vortex acoustics}}

\section{Introduction}

A great deal of work has been published lately based on the Unruh \cite{1} metric giving rise to the so-called analog gravity models \cite{2}. More recently Furtado et al \cite{3} have shown that non-Riemannian spaces can be useful to describe vortices in superfluids by investigating holonomy properties with torsion or the so-called translation holonomy. However, so far , no solid physical basis have appeared in the literature in order to obtain the so-called acoustic torsion which would play a analogous role of the Unruh metric which describes geometrically non-relativistic fluids in analog gravity. Recently work by Bergliaffa et al \cite{4} suggests that the Unruh metric would not be enough for describing fluids endowed with vorticity and other geometrical structure would be needed to investigate fluids with vorticity in the context of analog gravity. To fulfill this gap, in this paper we suggest that when the fluid is not irrotational, not only the Unruh metric is not enough to describe it but actually Cartan torsion \cite{5} is compulsory. Instead of using the Riemannian geometry of vortex acoustic in analogy to General Relativity (GR), here we investigate the non-relativistic rotational fluid equations by using the analogy with Einstein-Cartan gravity. According to Bilic \cite{6} extension to the relativistic acoustic geometry, one could generalize our results to obtain the relativistic fluids with acoustic torsion. Following these motivations, in this paper we provide two examples of distinct kinds of non-Riemannian gravity theories and their respective sonic analogs. Basically we consider two stationary torsion loops one in teleparallel gravity and the other in RC spacetime and their respectively sonic analogs. Their physical properties are distinct since in the first case it is shown that the fluid would possess vorticity proportional to torsion vector while in the second, the fluid would be irrotational. The paper is organized as follows: In the section 2 we present the wave equation with torsion and present the concept of acoustic torsion in  irrotationally perturbed non-relativistic fluid. In section 3 teleparallel torsion loop geometry first given by Letelier \cite{7} is reviewed. In section 4 torsion vector in terms of the superfluid vorticity is computed. We also show that the torsion loop line element can be conformally mapped to the acoustic line element and that the torsion vector also in Bose-Einstein (BEC) condensate is proportionally to the number density of vortices and to the vorticity vector. In section 5 we briefly show that teleparallel torsion loops maybe extended to general RC spacetime, which allows us to show that in this case the fluid is irrotational. Section 6 contains the conclusions. 
\section{Wave Equation in Torsioned Spacetime}
The starting point for the analogous models in GR was given by showing that the conservation equations for the fluid could be written in the Riemannian space if a proper effective metric is given. The fluid equations are expressed as the Riemannian wave equation for a scalar field ${\Psi}$ in the form
\begin{equation}
{\Delta}^{Riem}{\Psi}= 0
\end{equation}
where ${\Delta}^{Riem}$ represents the Riemannian D'Lambertian operator given by
\begin{equation}
{\Delta}^{Riem}=\frac{1}{\sqrt{-g}}{\partial}_{i}(\sqrt{-g}g^{ij}{\partial}_{j})
\end{equation}
Here $(i,j=0,1,2,3)$ and g represents the determinant of the effective Lorentzian metric which components are $g_{00}=-\frac{\rho}{c}(v^{2}-c^{2})$,$g_{0i}=-\frac{\rho}{c}(v_{i})$, $g_{11}=g_{22}=g_{33}= 1$, others zero. Throughout this Letter c represents the speed of sound,${\rho}$ is the fluid density, and $\vec{v}$ is the velocity of the fluid. Let us now apply the minimal coupling of torsion to the metric given by the covariant derivative of an arbitrary vector field $B_{k}$ in RC spacetime 
\begin{equation}
{\nabla}_{i}B_{j}={\partial}_{j}B_{j}- {{\Gamma}_{ij}}^{k}B_{k}
\end{equation}
where ${\Gamma}$ is the RC spacetime connection given in terms of the Riemannian-Christoffel connection ${\Gamma}'$ by 
\begin{equation}
{{\Gamma}_{ij}}^{k}={{{\Gamma}'}_{ij}}^{k}-{K_{ij}}^{k}
\end{equation}
where ${K_{ij}}^{k}$ are the components of the Cartan contortion. From these formulas we are able to express the non-Riemannian D'Lambertian as
\begin{equation}
{\nabla}_{i}{\Psi}^{i}={{\nabla}^{Riem}}_{i}{\Psi}^{i}+g^{ij}{K_{ij}}^{k}{\Psi}_{k}
\end{equation} 
To simplify future computations we consider the trace of contortion as given by $g^{ij}{K_{ij}}^{k}:=K^{k}$. We also define ${\Psi}^{i}={\partial}_{i}{\Psi}$. This definition allows us to express the non-Riemannian D'Lambertian of a scalar function as 
\begin{equation}
{\Delta}{\Psi}={{\Delta}^{Riem}}{\Psi}+ K^{k}{\partial}_{k}{\Psi}
\end{equation}
Thus the non-Riemannian wave equation  
\begin{equation}
{\Delta}{\Psi}=0
\end{equation}
reduces to the following equation
\begin{equation}
{{\Delta}^{Riem}}{\Psi}= - K^{k}{\partial}_{k}{\Psi}
\end{equation}
Let us now consider the dynamics of the non-relativistic fluids by writing the conservation of mass equation and the Euler equation as 
\begin{equation}
{\partial}_{t}{\rho}+ {\nabla}.({\rho}\vec{v})=0
\end{equation}
\begin{equation}
{\rho}(\frac{\partial\vec{v}}{{\partial}t})= {\rho}\vec{v}{\times}{\vec{\Omega}}-{\nabla}p -{\rho}{\nabla}({\phi}+\frac{1}{2}{\vec{v}}^{2})
\end{equation}
where ${\phi}$ is the potential energy of the fluid and p is the pressure. Here ${\vec{\Omega}}= {\nabla}{\times}{\vec{v}}$ is the vorticity of the fluid. Let us now perform the perturbation method to these equations according to the perturbed flow quantities 
\begin{equation}
{\vec{v}}= {\vec{v}}_{0}+ {\epsilon}{\vec{v}}_{1}
\end{equation}
\begin{equation}
{\rho}= {\rho}_{0}+ {\epsilon}{\rho}_{1}
\end{equation}
\begin{equation}
\vec{\Omega}= \vec{{\Omega}_{0}}+ {\epsilon}\vec{{\Omega}_{1}}
\end{equation}
Here we consider that only the velocity $\vec{v_{1}}$ is expressed as the gradient of a scalar field ${\Psi}_{1}$ since the unperturbed fluid is rotational, $\vec{v_{0}}$ cannot be expressed as the gradient of a scalar function ${\Psi}_{0}$, otherwise the fluid would be irrotational or in other words the vorticity $\vec{{\Omega}_{0}}$ would vanish. Thus only $\vec{{\Omega}_{1}}$ would vanish since
\begin{equation}
\vec{{\Omega}_{1}}={\nabla}{\times}{\vec{v_{1}}}={\nabla}{\times}{\nabla}{\Psi}_{1}=0
\end{equation}
The $0-index$ refer to the unperturbed quantities, while the prime indexed quantities are the perturbed ones, ${\epsilon}$ being a very small parameter. Substitution of these expressions into the conservation  equation yields
\begin{equation}
{\partial}_{t}{{\rho}_{1}}+{\nabla}.({\rho}_{1}\vec{v_{0}}+{\rho}_{0}\vec{v_{1}})=0
\end{equation}
using ${\xi}={\xi}_{0}+{\epsilon}\frac{p_{1}}{{\rho}_{0}}$ where ${\xi}=\frac{1}{\rho}{\nabla}p$, the Euler equation is
\begin{equation}
-{\partial}_{t}{\Psi}_{1}+\frac{p_{1}}{{\rho}_{0}}-\vec{{v}_{0}}.{\nabla}{\Psi}_{1}-{\alpha}=0
\end{equation}
where ${\nabla}{\alpha}:= -\vec{{\Omega}}_{0}{\times}{{\nabla}{\Psi}_{1}}$. Rearranging terms in the Euler equation
\begin{equation}
p_{1}={\rho}_{0}[{\partial}_{t}{\Psi}_{1}+\frac{p_{1}}{{\rho}_{1}}+\vec{v}_{0}.{\Psi}+{\nabla}{\alpha}]
\end{equation}
Substitution of $p_{1}$ into the conservation equation yields
\begin{equation}
-{\partial}_{t}(\frac{{\partial}{\rho}}{{\partial}p}{\rho}_{0}[{\partial}_{t}{\Psi}_{1}+\vec{v}_{0}.{\nabla}{\Psi}_{1}+{\alpha}])+{\nabla}.[{\rho}_{0}({\nabla}{\Psi}_{1}-\frac{{\partial}{\rho}}{{\partial}p}\vec{v_{0}}({\partial}_{t}{\Psi}_{1}+\vec{v}_{0}.{\nabla}{\Psi}_{1}+{\alpha}))]=0
\end{equation}
By comparison of this equation with the RC wave equation we note that the Riemannian wave part reproduces Unruh metric while the new part yields the following components of Cartan contortion
\begin{equation}
K^{0}\vec{v}_{1}= \frac{{\rho}_{0}}{c^{2}}\vec{\Omega}_{0}{\times}\vec{v}_{1}
\end{equation}
\begin{equation}
\vec{K}= \frac{{\rho}_{0}}{c^{2}}\vec{\Omega}_{0}{\times}\vec{v}_{0}
\end{equation}
where we have used $\frac{{\partial}{\rho}_{0}}{{\partial}p}=\frac{1}{c^{2}}$. One may note that by performing the scalar vector product of the equation for $K_{0}$ with $\vec{v_{1}}$, the zero component of contortion trace vanishes. This is not a problem because we always may choose a frame where $K^{0}$ vanishes. However by transvecting the $\vec{v_{0}}$ with the equation for the contortion vector $\vec{K}$ one obtains  $\vec{K}.\vec{v_{0}}=0$ , which shows that the contortion vector is orthogonal to the unperturbed initial flow given by ${\vec{v}_{0}}$. The same reasoning maybe used to show that the contortion vector $\vec{K}$ is also orthogonal to the vorticity $\vec{\Omega}_{0}$. This is particular interesting for two dimensional Helium II fluid.  
\section{Torsion loops in teleparallelism}
The solution representing a teleparallel torsion loop \cite{7} is given by the metric 
\begin{equation}
ds^{2}= (dt+\vec{B}.d\vec{r})^{2}-(dx^{2}+ dy^{2}+dz^{2})
\end{equation}
where $\vec{B}=(B_{x},B_{y},B_{z})$ is an arbitrary vector and $\vec{r}=(x,y,z)$. This metric can be written in terms of the basis 1-form ${\omega}^{i}$ where $(i,j=0,1,2,3)$ as
\begin{equation}
ds^{2}=({\omega}^{0})^{2}-({\omega}^{0})^{1}-({\omega}^{0})^{2}-({\omega}^{3})^{2}
\end{equation}
where ${\omega}^{0}=(dt+\vec{B}.d\vec{r})$, ${\omega}^{a}= dx^{a}$, and $(a,b=1,2,3)$. Let us now consider that torsion form is given by
\begin{equation}
T^{0}=  {\epsilon}_{abc}J_{a}{\omega}^{b}{\wedge}{\omega}^{c}
\end{equation}
Here ${\epsilon}_{abc}$ is the $3-D$ Levi-Civita totally skew symbol. Other torsion components $T^{a}=0$. From the Cartan's second structure equation 
\begin{equation}
{T^{i}}= d{\omega}^{i}+ {\omega}^{i}_{j}{\wedge}{\omega}^{j}
\end{equation}
one obtains the following \cite{7} nonvanishing connection one-form  
\begin{equation}
2{{\omega}^{i}}_{k}=-{{{{\epsilon}^{i}}_{k}}^{m}}_{n}{[{rot\vec{B}}-\vec{J}]}_{m}{\omega}^{n}
\end{equation}
Note that as pointed out by Letelier the relation
\begin{equation}
\vec{J}={\nabla}{\times}\vec{B}
\end{equation}
from the above connections would yield that all connection one-forms ${{\omega}^{i}}_{k}$ would vanish leaving behind a $T_{4}$ teleparallel spacetime where the full RC curvature would vanish from the first Cartan's structure equation 
\begin{equation}
{R^{i}}_{j}=\frac{1}{2}{R^{i}}_{jkl}{\omega}^{k}{\wedge}{\omega}^{l}=d{{\omega}^{i}}_{j}+ {{\omega}^{i}}_{p}{\wedge}{{\omega}^{p}}_{j}
\end{equation}
Letelier \cite{7} also showed that teleparallel torsion loops could be produced by choosing a torsion form as a distribution with support along a curve C with parametric equation ${\vec{x'}}={\vec{x'}({\lambda})}$ and taking the torsion vector ${\vec{J}}$ as 
\begin{equation}
{\vec{J}}(\vec{x})=I\int{{\delta}^{3}(\vec{x}-\vec{x'}({\lambda}))\frac{d{\vec{x}}}{d{\lambda}}d{\lambda}}
\end{equation}
where I is an arbitrary constant. The $\vec{J}$ components can be written as 
\begin{equation}
{\vec{J}_{k}(x)}=I {{\delta}_{k}}^{2}(\vec{x},C)
\end{equation}
This is a two-dimensional distribution with support on the line C. The integrability condition of the curl free for $\vec{B}$ expression is ${\nabla}.\vec{J}=0$. Application of above integrability condition to Dirac distribution for torsion $\vec{J}$ expression one obtains
\begin{equation}
{\nabla.\vec{J}}(\vec{x})={\delta}^{3}(\vec{x}-\vec{x'}_{i})-{\delta}^{3}(\vec{x}-\vec{x'}_{f})
\end{equation}
Thus in order that the torsion vector $\vec{J}$ be divergent-free one must impose that the final and initial points coincides or $\vec{x'}_{i}=\vec{x'}_{f}$. Therefore the Letelier metric has vanishing RC curvature and non-vanishing torsion along the curve C of the usual space. Unfortunately since in the teleparallel \cite{8} case the connection one-forms should vanish, we are not able to compute the holonomy and loop variables for the Letelier metric. 
\section{Acoustic Teleparallel Loops in BEC}
In this section we apply torsion loops in BEC \cite{9}. Let us now consider the dynamics of phonons propagating in the velocity field of the quantized vortex in Bose superfluid $He^{4}$ determined by the line element 
\begin{equation}
ds^{2}= (1-\frac{{v_{s}}^{2}}{c^{2}})(dt+\frac{Nkd{\phi}}{2{\pi}(c^{2}-{v_{s}}^{2})})^{2}-\frac{1}{c^{2}}(dr^{2}+ r^{2}d{\phi}^{2}+dz^{2})
\end{equation}
where $\vec{v_{s}}=\frac{Nk{\vec{e_{\phi}}}}{2{\pi}r}$ represents the velocity field around the quantized vortex,k is the quantum of circulation and N is the circulation quantum number. Here we shall make use of the Volovik approximation , $\frac{{v_{s}}^{2}}{c^{2}}<<1$ to obtain the cosmic spinning metric. Comparison between Unruh and Letelier metrics shows that they can be mapped into one another by the following coordinate transformations
\begin{equation}
dt'= \frac{1}{c}\sqrt{(1-\frac{{v_{s}}^{2}}{c^{2}})}dt
\end{equation}
and $r'=\frac{r}{c}$, ${\phi}'= {\phi}$, $z'=\frac{1}{c}z$. These expressions are the sonic analogs of relativistic Lorentz transformations. In order to complete the analogy we must add to these transformations the following expression
\begin{equation}
\vec{B}= \frac{Nk}{2{\pi}r\sqrt{(1-\frac{{v_{s}}^{2}}{c^{2}})}}\vec{e_{\phi}}
\end{equation}
The last expression allows us to compute the torsion vector $\vec{J}$ as
\begin{equation}
\vec{J}= -\frac{N^{3}k^{3}}{8{\pi}^{3}c^{2}r^{3}}{\vec{e_{z}}}
\end{equation}
We note here that the torsion vector is along the $z-direction$ and on the negative side of the vortex axis. It is also interesting to compare the last expression with the analogous expression for curvature given by Fischer and Visser \cite{10} 
\begin{equation}
R^{vortex}= \frac{2{r_{c}}^{2}}{r^{4}}
\end{equation}
which is the Ricci curvature scalar and $r_{c}$ is the "core radius".
In the next section we shall generalize the concept of torsion loops to RC spacetime and show that as far as our choice of spacetime structure is concerned it does not favor the formation of superfluid vortices \cite{11}. This constraint can now be applied to the superfluid metric to yield
\begin{equation}
\vec{\nabla}{\times}\vec{B}= -\frac{N^{3}k^{3}}{8{\pi}^{3}c^{2}r^{3}}{\vec{e_{z}}}
\end{equation}
Since this curl vanishes N is also zero, and the formation of superfluid vortices would not be favour in this geometry. From these last expressions it is easy to show that there is a connection between the superfluid vorticity $\vec{\Omega}_{s}:= {\nabla}{\times}\vec{v_{s}}$ and torsion vector $\vec{J}$ according to the expression
\begin{equation}
\vec{\nabla}{\times}\vec{v_{s}}= -\frac{N^{2}k^{2}}{2{\pi}r^{2}}{\vec{e_{z}}}
\end{equation}
which from the expression of $\vec{J}$ allows us to derive the following relation between the superfluid vorticity and torsion vector
\begin{equation}
\vec{J}= -\frac{N^{2}k^{2}}{4{\pi}c^{2}}{\vec{{\Omega}_{s}}}
\end{equation}
This shows that the result of section 2 which represents the connection between fluid vorticity and torsion vector ,also applies to the superfluid case. 
\section{Superfluids vortices as torsion loops in RC spacetime?} 
It is not difficult to note from section 2 that there is no need to fulfill condition $\nabla{\times}\vec{B}=0$ of the curl of ${\vec{B}}$ to have the integrability condition ${\nabla}.\vec{J}=0$ that leads to the construction of torsion loops. Actually this condition is equivalent to teleparallel spaces, but there is no need to impose teleparallel condition to have torsion loops. In fact in this section we propose a particular case of Letelier metric in the sense that now
\begin{equation}
{\nabla}{\times}\vec{B}=0
\end{equation}
This does not mean at all that the torsion vector vanishes, because we are imposing the "constraint" of teleparallelism. Yet indeed we may impose instead the constraint ${\nabla}.{\vec{J}}=0$ as a sort of torsion vector current conservation, which grant the fulfillement of conditions that lead to torsion loops this, time on a more general RC setting. Actually the torsion current conservation leads to torsion loops and this is not restricted only to teleparallelism. Therefore we may consider that the RC connection forms ${{\omega}^{i}}_{k}$ could be simply obtained from the respective teleparallel connections by simply taking $rot\vec{B}=0$. The resultant expressions could be used to compute the curvature forms from the first Cartan equation as 
\begin{equation}
{R^{0}}_{3}=\frac{1}{4}[({J_{x}}^{2}+{J_{y}}^{2}){\omega}^{3}-J_{y}J_{z}{\omega}^{2}]{\wedge}{\omega}^{0}
\end{equation}
\begin{equation}
{R^{3}}_{1}=-\frac{1}{2}[{{J_{y}},}_{y}{\omega}^{2}+{{J_{y}},}_{x}{\omega}^{1}+{{J_{y}},}_{z}{\omega}^{3}]{\wedge}{\omega}^{0}
\end{equation}
\begin{equation}
{R^{0}}_{2}=\frac{1}{4}[(2({{J_{z}},}_{z}+{{J_{x}},}_{x}){\omega}^{3}+{{J_{z}},}_{y}{\omega}^{2}){\wedge}{\omega}^{1}-2{{J_{x}},}_{y}{\omega}^{2}{\wedge}{\omega}^{3}-({{J_{z}}}^{2}{\omega}^{2}+J_{y}J_{z}{\omega}^{3}){\wedge}{\omega}^{0}]
\end{equation}
\begin{equation}
{R^{0}}_{1}=\frac{1}{4}[2({{J_{z}},}_{z}+{{J_{y}},}_{y}-{{J_{x}},}_{y}){\omega}^{3}{\wedge}{\omega}^{2}-({J_{z}}^{2}{\omega}^{1}-J_{x}J_{z}{\omega}^{3}){\wedge}{\omega}^{0}]
\end{equation}
\begin{equation}
{R^{2}}_{3}=-\frac{1}{4}[2({{J_{x}},}_{x}{\omega}^{1}+{{J_{x}},}_{y}{\omega}^{2}+{{J_{x}},}_{z}{\omega}^{3})]{\wedge}{\omega}^{0}
\end{equation}  
From these curvature expressions we note that the constraint ${\nabla}.{\vec{J}}=0$ only constraint the curvature expression but they do not vanish which caractherizes a nonteleparallel spacetime but indeed a general RC space. Since the spacetime curvature is purely torsionic off the loop, in our case, a test particle can feel torsion contrary to what happens in Letelier's torsion loops case. Therefore we may conclude that torsion loops may also be built in RC spaces. The vanishing of the curl of $\vec{B}$ also imposes that the metric of spacetime now has an axion type since $\vec{B}={\nabla}{\Phi}$ where ${\Phi}$ is a scalar field. 
\section{Conclusions}
We showed that acoustic superfluid metrics can be conformally related or rescaled to stationary torsion loop metrics. It is interesting to note that even with stationary torsion loops we are able to favour the formation of quantized superfluid vortices in the case of teleparallel spacetime, implying the existence of acoustic torsion. The same does not happen when we consider a special case of RC spacetime. Torsion loops may then be considered as torsion vortices inside the pulsars inner crusts of superfluid neutron star. The high density of neutron vortices may induce torsion loops and consequently the presence of torsion could be indirectly tested inside neutron stars \cite{12}. It is also important to note that the torsion loops discussed here could be used as analog geometrical models for the superconducting string loops \cite{13}.

\section*{Acknowledgements}
\paragraph*{}
I am very much indebt to Professor W. Unruh and Dr. S. Bergliaffa for helpful discussions on the subject of this paper. Thanks are also due to Prof. G. Volovik for pointed out a misprint in the first draft of this paper. Special thanks go to an unonymous referee for useful suggestions to improve the manuscript. Grants from CNPq (Ministry of Science of Brazilian Government) and Universidade do Estado do Rio de Janeiro (UERJ) are gratefully acknowledged.

\end{document}